\documentclass[aps,pra,superscriptaddress,floatfix,showpacs,10pt,twocolumn]{revtex4}
\usepackage[dvips]{graphicx}
\usepackage{amsmath}
\begin{document}
\title[Scheme for deterministic Bell-state-measurement-free quantum teleportation]{Scheme for deterministic Bell-state-measurement-free quantum teleportation}
\author{Ming Yang}
\email{mingyang@ahu.edu.cn}
\author{Zhuo-Liang Cao}
\email{zlcao@ahu.edu.cn} \affiliation{Anhui Key Laboratory of
Information Material {\&} Devices,School of Physics {\&} Material
Science, \\Anhui University, Hefei, 230039, PRChina}

\begin{abstract}
A deterministic teleportation scheme for unknown atomic states is
proposed in cavity QED. The Bell state measurement is not needed
in the teleportation process, and the success probability can
reach 1.0. In addition, the current scheme is insensitive to the
cavity decay and thermal field.
\end{abstract}

\maketitle

Quantum teleportation is one of the most important applications of
quantum entanglement. In quantum teleportation process, unknown
quantum information can be transmitted from a sender to a receiver
without the transmission of the carrier of the quantum
information\cite{bennett}. In the original scheme for quantum
teleportation, the sender and the receiver share a maximally
entangled state quantum channel. To realize the teleportation, the
sender must operate a joint Bell state measurement on the particle
that carries the unknown quantum information and one of the
entangled particles he possesses. Then the sender will inform the
receiver his measurement result. Finally, the receiver will
operate a single particle transformation on the particle he
possesses to transform the state into the unknown state to be
teleported based on the send's measurement result.

Recently, there is rapid progress in quantum teleportation of
unknown photon states. D.Bouwmeester \textit{et al} proposed the
first experimental demonstration for the teleportation of unknown
polarization photon states\cite{Bouwmeester}. Then the realization
of the deterministic\ teleportation of unknown continuous quantum
variables states of a beam of light was reported\cite{Furusawa}.
Pan \textit{et al} proposed the experimental realization of freely
propagating teleported qubits\cite{pan}.

We find that teleportation has been demonstrated in the domain of
photon. But, for the teleportation of unknown atomic states, it is
not the case. In cavity QED, several schemes have been proposed
for the teleportation of unknown atomic states. Cirac \textit{et
al} have made a proposal for quantum teleportation of an atomic
state by using two additional atomic levels of one of the
correlated pair\cite{cirac}. Bose \textit{et al} proposed a unique
proposal for the teleportation of atomic states via cavity
decay\cite{bose}. Some other teleportation schemes have been
proposed in cavity QED\cite{Davidovich, liwanli, me, me1, zheng}.
A more feasible scheme for quantum information processing has been
proposed\cite{zheng1}, where quantum teleportation of unknown
atomic state is realized by the dispersive interaction between two
atoms and a single-mode cavity. Because there is no exchange of
energy between atoms and cavity mode during the interaction, the
scheme is insensitive to cavity decay. Recently, superb progress
has been made in the realization of teleportation of atomic
states. Riebe \textit{et al} and Barrett \textit{et al }have
experimentally implemented the teleportation of the states of
Ca$^{+} $\cite{Riebe} and Be$^{+}$\cite{Barrett} respectively.

From analysis, we find that the key steps in quantum teleportation
are generation of quantum channel and realization of the join Bell
state measurement. Generation of entangled atomic states has been
implemented in experiment\cite{Osnaghi}. But the joint Bell state
measurement for atomic states is still a bottle-neck for the
quantum teleportation. Although the joint operation has been
realized in Refs\cite{Barrett, Riebe}, the operations needed there
are so complex. To overcome the difficulty of Bell state
measurement, Zheng proposed a teleportation scheme in cavity QED,
where the joint Bell state measurement is not needed\cite{zheng2}.
So the scheme is more feasible than the others. But, in the
scheme, the interaction between atoms and cavity is a resonant
one, and the cavity decay and the thermal field will affect the
scheme strongly. So Ye and Guo proposed another teleportation
scheme without Bell state measurement\cite{ye}. In their scheme,
the interaction is a dispersive one\cite{zheng1}, so the effect of
cavity decay has been eliminated. But the effect of thermal field
still exists. The success probabilities of these two schemes are
$0.25$ and $0.5$ respectively, i.e. they are all probabilistic
schemes.

In this paper, following a recent contribution on entanglement
generation\cite{zheng3}, we propose a deterministic teleportation
scheme without Bell state measurement, from which the effects of
thermal field and cavity decay are all eliminated. The success
probability is 1.0.

Next we will discuss the teleportation process in more details.
The sender possesses an atom $1$ in unknown state:
\begin{equation}
|\Phi\rangle_{1}=\alpha|e\rangle_{1}+\beta|g\rangle_{1},
\label{unkownstate}
\end{equation}
where $|e\rangle$ and $|g\rangle$ are the excited state and ground
state of atoms. Before considering the teleportation process
\textit{per se}, we will consider the generation of the quantum
channel. Consider two atoms $2$, $3$, which are all prepared in
ground states initially. To generate two-atom maximally entangle
states, a single-mode cavity must be introduced. Atoms $2$, $3$
will be sent through the cavity. At the same time, the two atoms
are driven by a classical field. Then the interaction between
atoms and the single-mode cavity can be described as follow:
\begin{align}
H&=\omega_{0}\sum_{j=1}^{2}S_{z,j}+\omega_{a}a^{+}a\label{hamiltonian}+\sum_{j=1}^{2}[g\left( a^{+}S_{j}^{-}+aS_{j}^{+}\right)\nonumber\\
& +\Omega\left(
S_{j}^{+}e^{-i\omega_{d}t}+S_{j}^{-}e^{i\omega_{d}t}\right)],
\end{align}
where $\omega_{0}$, $\omega_{a}$ and $\omega_{d}$ are atomic
transition frequency ($e\leftrightarrow g$), cavity frequency and
the frequency of driving field respectively, $a^{+}$ and $a$ are
creation and annihilation operators for the cavity mode, $g$ is
the coupling constant between atoms and cavity mode,
$S_{j}^{-}=|g\rangle_{j}\langle e|,$ $S_{j}^{+}=|e\rangle
_{j}\langle g|,$ $S_{z,j}=\frac{1}{2}\left(  |e\rangle_{j}\langle
e|-|g\rangle_{j}\langle g|\right)  $ are atomic operators, and
$\Omega$ is the Rabi frequency of the classical field. We consider
the case $\omega _{0}=\omega_{d}$. In the interaction picture, the
evolution operator of the system is~\cite{zheng2}:
\begin{equation}
U\left(  t\right)  =e^{-iH_{0}t}e^{-iH_{eff}t},
\label{totalevolution}
\end{equation}
where $H_{0}=\sum_{j=1}^{2}\Omega\left(
S_{j}^{+}+S_{j}^{-}\right)  $, $H_{eff}$ is the effective
Hamiltonian. In the large detuning $\delta\gg \frac{g}{2}$and
strong driving field $2\Omega\gg\delta,g$ limit, the effective
Hamiltonian for this interaction can be described as
follow~\cite{zheng2}:
\begin{align}
H_{eff}& =\lambda[\frac{1}{2}\sum_{j=1}^{2}(|e\rangle _{j}\langle
e|+|g\rangle_{j}\langle g|)
\label{effectivehamiltonian}\nonumber\\
& +\sum_{j,k=1,j\neq k}^{2}( S_{j}^{+}S_{k}^{+}+S_{j}^{+}
S_{k}^{-}+\text{H.c.})],
\end{align}
where $\lambda=\frac{g^{2}}{2\delta}$ with $\delta$ being the
detuning between atomic transition frequency $\omega_{0}$\ and
cavity frequency $\omega_{a}$. From the form of the effective
Hamiltonian, we conclude that the interaction Hamiltonian is
independent of the photon number of the cavity field. In addition,
there is no exchange of energy between atoms and cavity mode. So
the the effects of cavity decay and thermal field are all
eliminated.

After interaction time $t$, the state of the two atoms $2$, $3$
will undergo the following evolution:
\begin{align}
|g\rangle _{2}|g\rangle _{3}
\label{evolutioninitial}\longrightarrow& e^{-i\lambda t}[ \cos
\lambda t\left( \cos \Omega
t|g\rangle _{2}-i\sin \Omega t|e\rangle _{2}\right)\nonumber\\
& \times \left( \cos \Omega t|g\rangle _{3}-i\sin \Omega
t|e\rangle
_{3}\right)   \notag \nonumber\\
& -i\sin \lambda t\left( \cos \Omega t|e\rangle _{2}-i\sin \Omega
t|g\rangle _{2}\right)\nonumber\\
&\times \left( \cos \Omega t|e\rangle _{3}-i\sin \Omega t|g\rangle
_{3}\right)].
\end{align}
Choosing the interaction time $t$, we can let $\lambda
t=\frac{\pi}{4}$. Then the condition $\Omega t=\pi$ can be
realized by modulating the driving field appropriately. Then the
atoms $2$, $3$ will be left in:
\begin{equation}
|\Phi\rangle_{23}=\frac{1}{\sqrt{2}}\left(  |e\rangle_{2}|e\rangle
_{3}+i|g\rangle_{2}|g\rangle_{3}\right)  , \label{channel}
\end{equation}
which will be used as quantum channel. Suppose that the atoms $2$,
$3$ have been distributed among the sender (Alice) and the
receiver (Bob). To teleport the state of atom $1$ to atom $3$,
Alice will send the two atoms $1$, $2$ into a single-mode cavity.
At the same time, the two atoms $1$, $2$ are driven by a classical
field. This interaction has been described in Equations
(\ref{hamiltonian}, \ref{effectivehamiltonian}). The evolution of
the total system can be expressed as:
\begin{widetext}
\begin{align}
&  \left(  \alpha|e\rangle_{1}+\beta|g\rangle_{1}\right)  \otimes\frac
{1}{\sqrt{2}}\left(  |e\rangle_{2}|e\rangle_{3}+i|g\rangle_{2}|g\rangle
_{3}\right) \label{evolution}\nonumber\\
&  \longrightarrow\frac{1}{\sqrt{2}}\alpha e^{-i\lambda
t^{^{\prime}}}\left[ \cos\lambda t^{^{\prime}}\left(  \cos\Omega
t^{^{\prime}}|e\rangle_{1}-i\sin\Omega
t^{^{\prime}}|g\rangle_{1}\right)  \times\left(  \cos\Omega
t^{^{\prime}
}|e\rangle_{2}-i\sin\Omega t^{^{\prime}}|g\rangle_{2}\right)  \right. \nonumber\\
&  \left.  -i\sin\lambda t^{^{\prime}}\left(  \cos\Omega t^{^{\prime}%
}|g\rangle_{1}-i\sin\Omega t^{^{\prime}}|e\rangle_{1}\right)  \times\left(
\cos\Omega t^{^{\prime}}|g\rangle_{2}-i\sin\Omega t^{^{\prime}}|e\rangle
_{2}\right)  \right]  |e\rangle_{3}\nonumber\\
&  +\frac{i}{\sqrt{2}}\alpha e^{-i\lambda t^{^{\prime}}}\left[
\cos\lambda t^{^{\prime}}\left(  \cos\Omega
t^{^{\prime}}|e\rangle_{1}-i\sin\Omega
t^{^{\prime}}|g\rangle_{1}\right)  \times\left(  \cos\Omega
t^{^{\prime} }|g\rangle_{2}-i\sin\Omega
t^{^{\prime}}|e\rangle_{2}\right)  \right.
\nonumber\\
&  \left.  -i\sin\lambda t^{^{\prime}}\left(  \cos\Omega
t^{^{\prime}}|g\rangle_{1} -i\sin\Omega
t^{^{\prime}}|e\rangle_{1}\right)  \times\left( \cos\Omega
t^{^{\prime}}|e\rangle_{2}-i\sin\Omega
t^{^{\prime}}|g\rangle_{2}\right)
\right]  |g\rangle_{3}\nonumber\\
&  +\frac{1}{\sqrt{2}}\beta e^{-i\lambda t^{^{\prime}}}\left[
\cos\lambda t^{^{\prime}}\left(  \cos\Omega
t^{^{\prime}}|g\rangle_{1}-i\sin\Omega
t^{^{\prime}}|e\rangle_{1}\right)  \times\left(  \cos\Omega
t^{^{\prime} }|e\rangle_{2}-i\sin\Omega
t^{^{\prime}}|g\rangle_{2}\right)  \right.
\nonumber\\
&  \left.  -i\sin\lambda t^{^{\prime}}\left(  \cos\Omega
t^{^{\prime} }|e\rangle_{1}-i\sin\Omega
t^{^{\prime}}|g\rangle_{1}\right)  \times\left( \cos\Omega
t^{^{\prime}}|g\rangle_{2}-i\sin\Omega t^{^{\prime}}|e\rangle
_{2}\right)  \right]  |e\rangle_{3}\nonumber\\
&  +\frac{i}{\sqrt{2}}\beta e^{-i\lambda t^{^{\prime}}}\left[
\cos\lambda t^{^{\prime}}\left(  \cos\Omega
t^{^{\prime}}|g\rangle_{1}-i\sin\Omega
t^{^{\prime}}|e\rangle_{1}\right)  \times\left(  \cos\Omega
t^{^{\prime} }|g\rangle_{2}-i\sin\Omega
t^{^{\prime}}|e\rangle_{2}\right)  \right.
\nonumber\\
&  \left.  -i\sin\lambda t^{^{\prime}}\left(  \cos\Omega
t^{^{\prime} }|e\rangle_{1}-i\sin\Omega
t^{^{\prime}}|g\rangle_{1}\right)  \times\left( \cos\Omega
t^{^{\prime}}|e\rangle_{2}-i\sin\Omega t^{^{\prime}}|g\rangle
_{2}\right)  \right]  |g\rangle_{3}.
\end{align}
\end{widetext} After the interaction time $t^{^{\prime}}$, Alice
will detect the atoms $1$, $2$. If the result is
$|e\rangle_{1}|e\rangle_{2}$, the state of atom $3$ will collapse
into the following state:
\begin{align}
|\Psi\rangle_{3}  &  =\frac{1}{\sqrt{2}}\alpha e^{-i\lambda
t^{^{\prime}} }\left(  e^{-i\lambda t^{^{\prime}}}\cos^{2}\Omega
t^{^{\prime}}+i\sin\lambda
t^{^{\prime}}\right)  |e\rangle_{3}\label{result}\nonumber\\
&  -\frac{i}{\sqrt{2}}\beta e^{-i\lambda t^{^{\prime}}}\left(
e^{-i\lambda t^{^{\prime}}}\sin^{2}\Omega
t^{^{\prime}}+i\sin\lambda t^{^{\prime}}\right)
|g\rangle_{3}\nonumber\\
&  +\frac{1}{2\sqrt{2}}\alpha e^{-2i\lambda t^{^{\prime}}}\sin2\Omega
t^{^{\prime}}|g\rangle_{3}\nonumber\\
&  -\frac{i}{2\sqrt{2}}\beta e^{-2i\lambda t^{^{\prime}}}\sin2\Omega
t^{^{\prime}}|e\rangle_{3}.
\end{align}
If the interaction time satisfies the condition $\lambda
t^{^{\prime}} =\frac{\pi}{4}$and $\Omega t^{^{\prime}}=\pi$, the
state in Equation (\ref{result}) becomes:
\begin{equation}
|\Psi^{^{\prime}}\rangle_{3}=\frac{1}{2}\left( \alpha|e\rangle_{3}
+\beta|g\rangle_{3}\right)  .\label{finalresult}
\end{equation}
So the teleportation succeeds with probability $\frac{1}{4}$. If
the measurement results are
$|g\rangle_{1}|g\rangle_{2},|g\rangle_{1}|e\rangle
_{2},|e\rangle_{1}|g\rangle_{2}$, the teleportation also can
succeed with probability $\frac{1}{4}$(As depicted in the
Table.$1$). That is to say the total success probability is $1.0$,
which is more higher than that of the previous ones. Table.$1$
lists the measurement results on atoms $1$, $2$, the result state
of atom $3$ and the operation needed for the receiver to convert
the state of atom $3$ into the initial state of atom $1$.
\begin{table}
 \caption{\label{Table.1.}The results of the teleportation scheme. M.R. denotes the measurement result on atoms 1, 2, Operation denotes the operation needed for the receiver.}
\begin{ruledtabular}
\begin{tabular}{lcr}
M.R.& $|\Psi\rangle_{3}$ & Operation\\
$|e\rangle_{1}|e\rangle_{2}$ & $\frac{1}{2}\left(
\alpha|e\rangle_{3}
+\beta|g\rangle_{3}\right)  $ & I\\
$|g\rangle_{1}|g\rangle_{2}$ & $\frac{1}{2}\left(
\alpha|e\rangle_{3}
-\beta|g\rangle_{3}\right)  $ & $\sigma_{z}$\\
$|e\rangle_{1}|g\rangle_{2}$ & $\frac{1}{2}\left(
\alpha|g\rangle_{3}
-\beta|e\rangle_{3}\right)  $ & $\sigma_{y}$\\
$|g\rangle_{1}|e\rangle_{2}$ & $\frac{1}{2}\left(
\alpha|g\rangle_{3} +\beta|e\rangle_{3}\right)  $ & $\sigma_{x}$
\end{tabular}
\end{ruledtabular}
\end{table}

Next, Consider the feasibility of the current scheme. Because the
scheme is insensitive to cavity decay and thermal field, the
scheme only must be completed within the radiative time of the
atoms. For Rydberg atoms with
principal quantum numbers $50$ and $51$, the radiative time is $T_{a}
=3\times10^{-2}s$. From the analysis in reference\cite{zheng1},
the interaction time is on the order $t\simeq2\times10^{-4}s$,
which is much shorter than the atomic radiative time $T_{a}$. So
our scheme is reliable by using cavity QED techniques.

In conclusion, we proposed a deterministic scheme for the
teleportation of unknown atomic states, where the Bell state
measurement is not needed. The distinct advantage of the current
scheme is that, not only teleportation of unknown atomic states
can be realized without the Bell state measurement, but also the
success probability can reach $1.0$ with fidelity being $1.0$. The
dispersive interaction between two driven atoms and a single-mode
cavity used here makes the current scheme possess another distinct
advantage, i.e. the effects of cavity decay and thermal field are
all eliminated.  But, there is still a disadvantage here. We must
distinguish the two atoms $1$, $2$ after they flying out of the
cavity. Fortunately, Riebe \textit{et al }\cite{Riebe} have
developed a technique to address any specified target ion using
tightly focused laser beams and to `hide' the remaining ions from
the target ion's fluorescence by changing their internal states so
that they are insensitive to the fluorescent light. So the scheme
is feasible in the current technology.

\begin{acknowledgements}
This work is supported by the Natural Science Foundation of the
Education Department of Anhui Province under Grant No: 2004kj005zd
and Anhui Provincial Natural Science Foundation under Grant No:
03042401 and the Talent Foundation of Anhui University.
\end{acknowledgements}

\end{document}